\documentclass[submission, Phys]{SciPost}
\usepackage{amsmath,amssymb,bm}
\usepackage{times}
\usepackage{float}
\usepackage{color}
\usepackage[]{hyperref}
\hypersetup{linkbordercolor=red}
\usepackage{soul}
\usepackage[normalem]{ulem}
\usepackage[english]{babel}
\usepackage[OT1]{fontenc}
\usepackage[capitalize]{cleveref}
\usepackage{epstopdf}
\usepackage{graphicx}
\DeclareGraphicsExtensions{.png,.eps}
\graphicspath{{fig/}}
\usepackage{braket}

\newcommand\be{\begin{equation}}
\newcommand\bea{\begin{eqnarray}}
\newcommand\bes{\begin{subequations}}
\newcommand\esu{\end{subequations}}
\newcommand\ee{\end{equation}}
\newcommand\emul{\end{multline}}
\newcommand\eea{\end{eqnarray}}

\usepackage{etoolbox}
\makeatletter
\patchcmd{\ttlh@hang}{\parindent\z@}{\parindent\z@\leavevmode}{}{}
\patchcmd{\ttlh@hang}{\noindent}{}{}{}
\makeatother

\begin{document}

\begin{center}{\Large \textbf{
Characterizing the quantum field theory vacuum using temporal Matrix Product 
States
}}\end{center}

\begin{center}
Emanuele Tirrito$^{1,\ast}$, Neil J. Robinson$^{2}$, Maciej Lewenstein$^{1,3}$,  Shi-Ju Ran$^{4,1}$, Luca~Tagliacozzo$^{5,\dagger}$
\end{center}

\begin{center}
{\bf1} ICFO-Institut de Ciencies Fotoniques, The Barcelona Institute of Science and Technology, 08860 Castelldefels (Barcelona), Spain\\
 {\bf 2} Institute for Theoretical Physics, University of Amsterdam, Postbus 94485, 1090 GL Amsterdam, The Netherlands\\
{\bf 3} ICREA, Passeig Lluis Companys 23, 08010 Barcelona, Spain \\
{\bf 4} Department of Physics, Capital Normal University, Beijing 100048, China \\
{\bf 5} Instituto de Física Fundamental IFF-CSIC, Calle Serrano 113b, Madrid 28006, Spain
\end{center}

\begin{center}
$^{\ast}$emanuele.tirrito@icfo.eu \qquad
$^{\dagger}$luca.tagliacozzo@iff.csic.es\\  
  \vspace{3mm}
\end{center}

\begin{center}
\today
\end{center}


\section*{Abstract}
{\bf
In this paper, we construct the continuous Matrix Product State (MPS) 
representation of the vacuum of the  field theory  corresponding to the 
continuous limit of an Ising model. We do this  by exploiting the observation made by Hastings and Mahajan in {\color{blue} Phys. Rev. A \textbf{91}, 032306 (2015)}  that the Euclidean time evolution generates a continuous MPS along the time direction. This fact, together with the emerging Lorentz invariance at the critical point allow to identify  the matrix product representation of the quantum field theory (QFT) vacuum with the continuous MPS in the time direction (tMPS). We explicitly construct the tMPS and check these statements by comparing the physical properties of the tMPS with those of the standard ground state MPS. We furthermore characterize the QFT that the tMPS encodes by identifying the spectrum of an Ising field theory using the Bisognano Wichmann theorem.}

\vspace{10pt}
\noindent\rule{\textwidth}{1pt}
\tableofcontents\thispagestyle{fancy}
\noindent\rule{\textwidth}{1pt}
\vspace{10pt}

\section{Introduction} \label{sec:intro}
Tensor networks can efficiently encode equilibrium states of quantum many-body systems described by local Hamiltonians \cite{B+68,B+78,OR+95,NO+96,VMC+08,V03,V04,DKS04,V07,OV08,JOV+08,OV+09,RO+11,RO+14,HCO+11}, as well as classical partition functions obtained from the quantum-classical correspondence. This was first realised in White's density matrix renormalization group algorithm that, as of today, remains the best numerical tool for characterising strongly correlated one-dimensional (1-D) quantum systems \cite{W+92,W+93,S+11,M+07,M+08}. Tensor network approaches were soon extended to higher dimensions and connected with theories of entanglement in quantum many-body systems being developed simultaneously in the quantum information community \cite{VLR+03,TOI+08,PMT+09,NOHMA+00,NOHMA+01,VC+04,LN07,JWX08,GLW08,ECP+09,CCD12,SW+12,PBT+15}. 

Recently, tensor networks have been applied to quantum field theories (QFTs). Rather than starting from a lattice model, as is standard with tensor network algorithms, once can instead start from a continuum Hamiltonian that describes a QFT and formulate a variational calculation that expresses the vacuum of the QFT as a tensor network \cite{VC+11,HCOV+13,HOVV+13,MHO+13,JBHO+15,BHVVV+16,DHVR+17,GRV+17,GV+18,TC+18,HFV+18,delcamp_2020}. This is rather different to the standard approach of taking the continuum limit, familiar in both condensed matter physics \cite{SS+11} and lattice field theories \cite{MM97}, where one reduces the spacing between sites in a lattice model. To be more precise, in a lattice model with lattice spacing $a$ and correlation length $\xi = \bar \xi a$ this entails taking two limits simultaneously, $a\to0$ and $\bar\xi\to\infty$, such that the correlation length $\xi = \bar \xi a$ is fixed to correctly reproduce the desired physics of the continous.
As a result, in order to describe a field theory, the $\bar \xi$ must diverge but only for masless field theories, $\xi$ also diverges in the continuum.

Taking such a scaling limit in a numerical simulation can, in practice, be tricky. In this work we will consider a particular limiting procedure that can simplify proceedings considerably if the model is sufficiently close to a  quantum critical point. The main idea here is as follows: Usually to find the ground state of a lattice model using tensor network algorithms, one performs imaginary time-evolution (initiated from a random initial state). Let us consider the problem of imaginary time-evolution on its own; the introduction of imaginary time transforms a $D$-dimensional lattice problem into a $(D+1)$-dimensional problem. When interested in the ground (i.e., zero temperature) state of a problem, imaginary time is a continuous variable, $\beta \in [0,\infty]$. In computations one has to perform imaginary time-evolution to a large but finite $\beta = \beta_f \gg 1$, and one does so by breaking the interval $[0,\beta_f]$ into many small steps $\Delta\beta \ll 1$ (i.e., one imposes a lattice in the imaginary time direction). For each step $\Delta\beta$ the action of the imaginary time-evolution operator is approximated using the Suzuki-Trotter decomposition \cite{S90,S91}. This approximation becomes more and more accurate as the step size $\Delta\beta$ is taken to zero, that is, as we take the continuum limit in the imaginary time direction (as observed by Hastings \cite{HM+15}). 

From this construction, one can extract a continuum MPS along the imaginary time direction, called the \textit{temporal MPS} (tMPS), which will be discussed in detail in the following section. At first glance, it is not obvious what the tMPS encodes and in what manner it is useful. In one special case, however, light can be shed on the situation. If the model being studied has an emergent Lorentz-invariant critical point, and the parameters of the model are tuned such that we are studying the critical point, then the tMPS encodes directly the continuum limit of the lattice ground state MPS (with the two being related via the sound velocity). In other words, in this special case the tMPS encodes the vacuum of the emergent QFT. In the general case, this is not true. 

There are some subtleties, however, that have been glossed over in the above descriptions:

\begin{enumerate}

\item The power of tensor network approaches comes from their ability to efficiently encode equilibrium quantum states in local models. They do so systematically truncating the information required to describe such a quantum state via the introduction of a finite bond dimension, $\chi$, which limits the size of the tensors constructed within the algorithms. Taking the continuum limit, this has the effect of limiting the number of degrees of freedom (per unit length), which has clear implications on the bipartite entanglement of the QFT vacuum (making it finite) \cite{TOI+08, VC+04}.  Similarly, finite bond dimension is understood to introduce a finite correlation length (and thus cannot exactly describe the physics at a quantum critical point) and can break discrete symmetries. However, the precise role played the bond dimension as a regulator within the QFT remains unclear \cite{TOI+08,NOK+96,ABO+99,SHM+15,LT+12,KPM+11,ZMV+17,CCKT+18,RL+18}. 

\item Starting from a lattice model, the low-energy spectrum of the emergent field theory depends on both the critical exponents at the quantum critical point and the boundary conditions of the lattice model \cite{Cardy+861, Cardy+862,evenbly+10}. In the case of an infinite system, the issue of boundary conditions is unclear; to get around this, one can introduce an impurity that tunes between all the possible conditions \cite{barev,mccoy_1980,oshikawa_1997}. We note that a connection between the physics of impurities and the structure of entanglement in MPS has been proposed in \cite{bal_2016}. 

\end{enumerate}

The first subtlety is essentially related to the order of limits. In a perfect world, one would like to first take the continuum limit $a\to0$ and then truncate the information $\chi\to{\rm const.}$. In practice, one must first truncate the information $\chi\to{\rm const.}$ and then take the continuum limit of this truncated representation, $a\to0$. It has already been observed, through finite size scaling calculations, that these two different procedures will generally generate different fixed-point tensors \cite{LT+12}. In other words, the two limits do not commute. 

With this work, we hope to shed some further light on these issues for the case of a simple and paradigmatic lattice model: the transverse field Ising chain. We will tune the lattice model to its quantum critical point, construct the trotterized-tMPS at finite bond dimension, and then take the continuum limit for the imaginary-time direction to obtain the tMPS. 

By exploting the well known correspondence between the spectrum of the reduced density matrix of half infinite system and the spectrum of the field theory, dictated by the Bisognano Wichmann theorem \cite{bisognano_duality_1975} we actually verify that the tMPS describes an Ising field theory. 

The spectrum of the reduced density matrix of the system is in general given by the spectrum of the field theory with open boudnary conditions, and it is not affected by the true boundaries of the system. 

In constructing the tMPS, we naturally obtain a transfer matrix.
As the tMPS transfer matrix propagates the dynamical correlations of the problem, naively one would expect to approximately recover the expected mass ratios of the emergent QFT, and in this case we should be sensitive to the precise boundary conditions. In the case at hand, we examine the transfer matrix and compare it to the known mass spectrum of the Ising field theory, and various perturbations of it. 

Unfortunately,we are unable to match the spectrum obtained from the tMPS to an Ising model with an impurity, or cases where the field theory is perturbed by the leading relevant operators of the theory. In particular, we are unable to reproduce the transfer matrix eigenvalues using Zamolodchikov's perturbed Ising model, which introduces both a mass for the fermions and a longitudinal (confining) magnetic field \cite{Cardy+862, Henkel+87, Zam-89}. 

We discuss possible subtelties of the identification and we leave as an open question to precisly understand the relation between the spectrum of the tMPS transfer matrix and the spectrum of the emergent QFT. 

This paper proceeds as follows. In the next section, we introduce some useful definitions and preliminaries for constructing tensor networks, including the temporal MPS. In section 3 and 4 we explicitly construct the tMPS and characterize the QFT that the tMPS encodes in terms of its spectrum. In section 5 a summary and an outlook are presented. 


\section{Basic definitions and preliminaries} 

\begin{figure}
    \begin{center}
	\includegraphics[width=\textwidth]{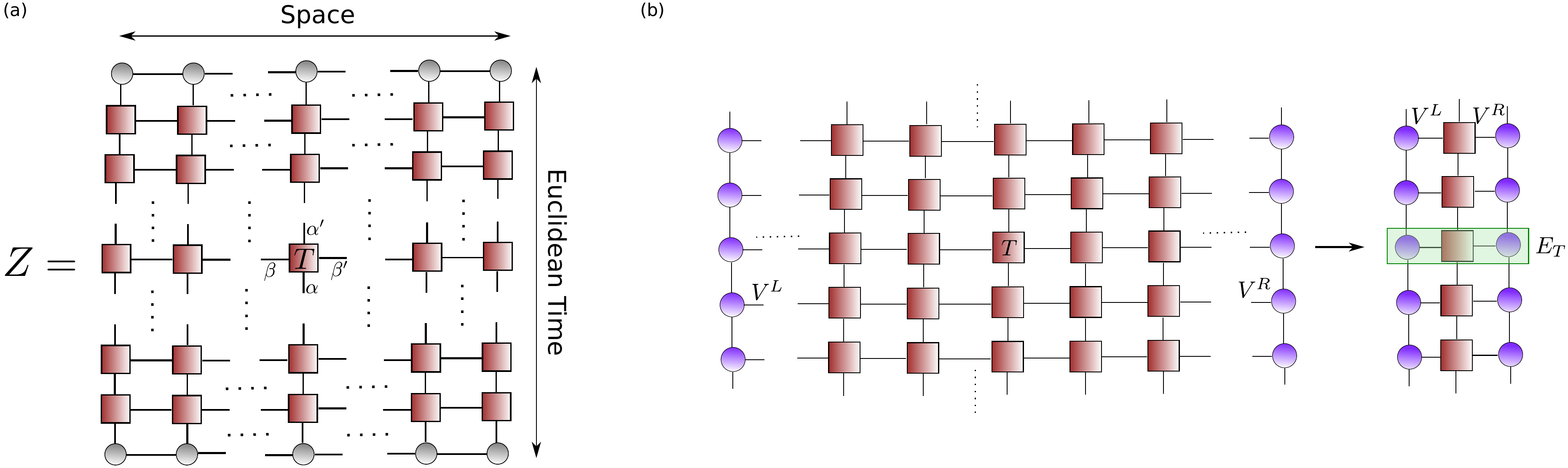}  
	\end{center}
	\caption{\textbf{2D tensor network and temporal matrix product 
	state.} (a) Representation of 2D tensor network encoding the 
	vacuum 	to vacuum transition probability corresponding to the 
	norm of the ground state obtained by performing an imaginary 
	time evolution of an initial matrix product state. Geometric 
	shapes represent the elementary tensors, and the lines encode 
	their contraction. (b)The MPS fixed point obtained  by 
	contracting the 2D TN from left to right defines the tMPS. It 
	describes a state along the Euclidean time direction and 
	becomes a continuous MPS in the limit of the Trotter step 
	going to zero.}
	\label{fig-prb}
\end{figure}

We want to  evaluate a 2D TN with the structure shown in Fig. \ref{fig-prb}(a). The TN will be contracted to a scalar $Z$ whose value could represent different physical quantities, depending on the content of the individual tensors. For example, $Z$ could encode the partition function of a  $2D$ classical model, the norm of a 2D quantum state encoded in a PEPS, or the real or imaginary time evolution of a 1D quantum state. In our paper, $Z$ will encode the latter case, the imaginary time evolution of a 1D quantum state $\vert \psi \rangle$.

The $1D$ system is made by  $N$ constituents whose interaction is described by, for simplicity, a nearest neighbors Hamiltonian $\hat{H}= \sum_n \hat{H}^{[n,n+1]}$. $\hat{H}^{[n,n+1]}$ encodes the two-body interactions. If a system is translationally invariant, as in the cases we will consider here all the two body terms are the same, namely $\hat{H}^{[n,n+1]} = \hat{H}^{[2]}$.

The imaginary time evolution, performed for sufficiently large times,  allows to approximate the ground state of $\hat{H}$ \footnote{We assume that $\braket{\Omega|\psi}\neq0$ as expected for a randomly chosen $\ket{\psi}$.},
\be \label{MinP-eq}
|\Omega \rangle = \lim_{\beta\rightarrow \infty} \frac{e^{-\frac{\beta 
}{2}\hat{H}} 
|\psi \rangle}{|| e^{-\frac{\beta}{2} \hat{H}} |\psi \rangle ||}.
\ee

Alternatively, the ground state  of $\hat{H}$ could be obtained  variationally by minimizing the energy over the class of normalized MPS states with fixed bond dimension,
\be 
|\Omega \rangle = \arg \min_{|\psi \rangle} \left\lbrace \langle \psi | \hat{H} | \psi \rangle \right\rbrace.
\ee
$|\psi \rangle$ is an MPS state fulfilling $\langle \psi | \psi \rangle=1$.

At large fixed $\beta$ in eq. \eqref{MinP-eq}  we need to divide the evolution in small steps by fixing $M$ such that $\beta/2M=\tau \ll 1$. In this way, we can approximate Eq. (\ref{MinP-eq}) step by step using a Suzuki-Trotter decomposition at the chosen order in $\tau$. 

When $|\psi \rangle$  is in an MPS form,  each step  $U(\tau)|\psi \rangle \equiv e^{-\tau \hat{H}} |\psi \rangle$ can be performed approximately  by first contracting the TN and then truncating the MPS back to desired bond dimension D after normalizing the state. The fixed point of this procedure provides the MPS representation of $|\Omega \rangle$. 

As a result, in the case of the imaginary time evolution, the individual tensors of the 2D TN represented in  Fig. \ref{fig-prb}, are related to the small steps of time evolution $U(\tau)$.

The length of horizontal direction encodes the number of constituents $N$ of the  $1D$ quantum system, and we call it in the following spatial direction. The vertical length encodes the number of Trotter steps $M$ and we call it  Euclidean-temporal direction. 

Looking at the $2D$ TN we can envisage a different contraction scheme (see Ref. \cite{BHV+09, HM+15}). Rather than contracting the TN downwards, we can contract it from left to right as in 
(Fig. \ref{fig-prb} (b)). Once again, if we enforce an MPS structure for the contraction (alternating contraction and truncation steps) we can define the tMPS as the fixed point of this contraction strategy,
\be
|tMPS\rangle = \sum_{\{\beta\}} \cdots V_{\alpha_i \beta_i \beta_{i+1}} V_{\alpha_j \beta_j \beta_{j+1}}\cdots |\cdots \alpha_i \alpha_j \cdots \rangle.
\ee
The dots encode the fact that the tMPS is infinite ($\beta$ and $M$ diverge), and $V$ denotes its constituent tensors. Notice that the quantum state is now defined as a state with a fixed ``auxiliary'' position and one constituent per Trotter step, thus effectively encoding the time evolution of a single coarse-grained constituent. 
In the next sections, we characterize the tMPS through the corner transfer matrix renormalization group algorithm (CTMRG) \cite{NO+96,NOHMA+00,NOHMA+01,NOK+96}.

\section{The continuous  tMPS}

\begin{figure}[t]
    \begin{center}   
	\includegraphics[width=\textwidth]{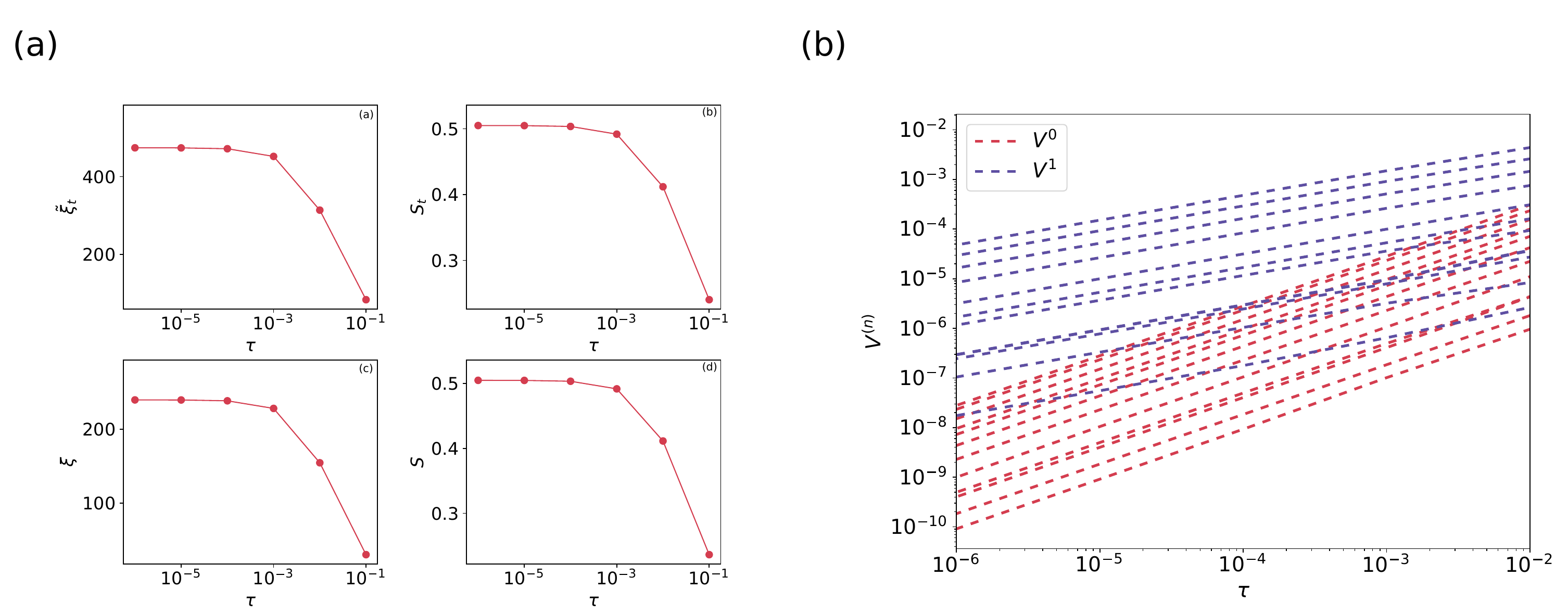}
	\end{center}
	\caption{\textbf{Convergence and tMPS elements.}
	($a_1$)The physical    
	correlation time $\tilde{\xi}_t$ versus $\tau$. We appreciate 
	that $\tilde{\xi}_t$ stays finite when $\tau$ goes to zero. 
	The system is at the critical point $h=J$ where we expect the 
	correlation time to  diverge. However, $\tilde{\xi}_t$ stays 
	finite as a result of the  finite bond dimension of the tMPS. 
	\textit{Entanglement entropy of the tMPS} ($a_2$). The 
	entanglement entropy of the tMPS only depends on the physical 
	correlation time $\tilde{\xi}_t$ and stays finite when $
	\tilde{\xi}_t$ is finite.  At the critical point, the 
	linear dispersion relation of the low-energy excitations 
	implies an enhanced space-time symmetry. This can be checked 
	by studying the correlation length in the ground state $\xi$ 
	($a_3$-$a_4$). Once more, it is finite as a result of the 
	finite bond dimension $\chi$.  The entanglement entropy $S$ of 
	half of the ground state also stays finite, and weakly depends 
	on $\tau$ as expected. We take the bond dimension cut-off 
	$\chi=20$.(b) By plotting the out-diagonal elements of the 
	tMPS tenors as a 
	function of $\tau$ we can identify the elements whose 
	extrapolation would produce $V^0$ and $V^1$ contribution in  
	[Eqs. (\ref{eq:cv0}) and (\ref{eq:cv1})]. They are 
	distinguishable  by the different slope of their $\tau$ 
	dependence. We could thus reconstruct the$V^0$ and $V^1$ from 
	a simple extrapolation of the finite $\tau$
	data. These plots are a further confirmation that  the tMPS
	converges to a continuous MPS.}
	\label{fig-ConvergeTau_CompV0V1}
\end{figure}

Besides very specific scenarios, the fixed points extracted after the contraction 
along the spatial and the temporal directions are different. The lattice spacing along the horizontal direction  is discrete, while, in the limit $M\to\infty$ it becomes continuous in  the vertical direction. While the parallel MPS represents the ground state of $\hat{H}$ the tMPS sites  encode the different instant of time of the  evolution of a single coarse-grained constituent. Since in the limit $M\to\infty$ the time step goes to zero, the constituent continuously varies in time, and thus the tMPS encodes a continuous system \cite{VC+11}.

When the $2D$ TN represents a classical isotropic model, the fixed points obtained by contracting the network along either the vertical or the horizontal directions should represent the same state.  This equivalence has been tested on the TN that encodes the partition function of the $2D$ classical Ising model  (see Appendix \ref{2DCIM}). This scenario can also occur in the quantum case in very special cases, where the original $\hat{H}$  possesses extra symmetries such as e.g, the emerging Lorentz invariance we will consider in the following. 

While the properties of the spatial MPS have been analyzed in several works, here we want to characterize the properties of the tMPS. This is a relatively unexplored area, and we only are aware of the results presented by Hasting and Mahajan \cite{HM+15}.

We characterize the continuous nature of the tMPS by addressing the 
paradigmatic  quantum Ising model defined by the Hamiltonian 
\begin{eqnarray} \label{eq-IsingHam}
H=J \sum_{i} S^x_i S^x_{i+1} - h\sum_{i} S^z_{i},
\end{eqnarray}
with $S^x$ and $S^z$ are the Pauli matrices along the $x$ and $z$ directions. 
The model can be exactly solved \cite{SML64, P70} . The system is in a 
paramagnetic phase for values of the couplings such that $h/J>1.0$. In 
the thermodynamic limit, the order parameter acquires a non-vanishing expectation value $\langle S^x \rangle \neq 0$ in the ferromagnetic phase. In the paramagnetic phase, for $h/J>1.0$ the order parameter goes to zero. The two phases are separated by a quantum phase transition at $h/J=1.0$, where the low-energy physics of the system can be described by a conformal field theory with central charge $c=0.5$ \cite{BPZ84,C96,G98}.

The continuous nature of the tMPS emerges when taking the limit $M\to \infty$. If the system in the continuum has a finite correlation time $\tilde{\xi}_T$, we expect that the correlation time measured in terms of the lattice spacing $\xi_T$ should diverge in the limit as  
$\xi_T =\tilde{\xi}_T /\tau$. This is the first check that we perform on our tMPS since the correlation time of a translationally invariant tMPS is encoded in the gap of its transfer matrix $E_T \equiv \sum_{\alpha} V_{\alpha}\bar{V}_{\alpha}$,
\be
\xi_T = \frac{1}{\ln \eta_0 - \ln \eta_1},
\label{eq-Xi}
\ee
where  $\eta_n$ represents the $n$-th eigenvalue of $E_T$. 

This property is analyzed in  Fig. \ref{fig-ConvergeTau_CompV0V1} (a). The correlation time $\xi_T$  obtained directly from the tMPS [Eq. (\ref{eq-Xi})] diverges as expected as  $\tau$ decreases. In panel 
$(a_1)$ we show that, as expected, the physical correlation time
$\tilde{\xi}_T$ converges rapidly when reducing $\tau$. In panel $(a_2)$ we check that a physical quantity, the entanglement entropy of half of the temporal chain $S$, only depends on $\tilde{\xi}_T$. $S$ should indeed scale as $S\propto \log{\tilde{\xi}_T}$ \cite{CC+04}.

Despite working at the critical point $h=J$ where the correlation time $\tilde{\xi}_T$ should diverge  since $\hat{H}$ is gap-less, we observe a finite correlation time. This is a consequence of describing the system with a finite bond dimension $\chi$, thus effectively cutting-off the correlations.

\begin{figure}[t]
    \begin{center}
	\includegraphics[width=\textwidth]{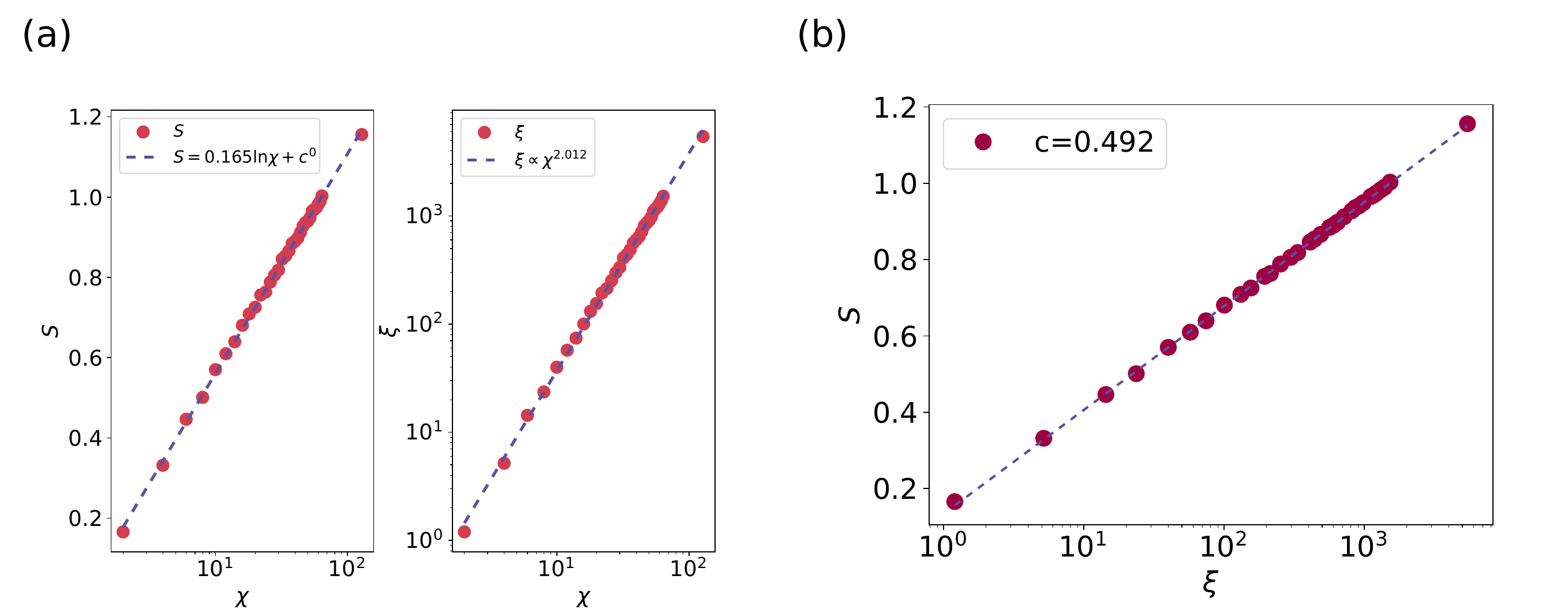}
	\end{center}
	\caption{\textbf{Entanglement entropy and correlation length.}
	(a) As a sanity check, we reproduce the well-known scaling of 
	the correlation length $\xi$ and entanglement entropy $S$ 
	versus the bond dimension cut-off $\chi$ of the spatial MPS. 
	We take $h=1$, and $\tau=10^{-6}$. 
	(b) The half chain entanglement entropy $S$  as a function of 
	the  correlation length $\xi$ at the critical point. Each 
	point is obtained by choosing the bond dimension $\chi$. A fit 
	to the expected logarithmic scaling allows us to extract  
	$c=0.5$ as expected.}
	\label{fig-Q1D_spatial}
\end{figure}

At the critical point, at low energies, thanks to the linear dispersion relation of the low-energy excitations \cite{P70}, we expect to observe an enhanced symmetry between space and time. Indeed, momentum plays the same role than energy, and the theory becomes Lorentz invariant. 

In panel $(a_3)$ and $(a_4)$ of Fig. \ref{fig-ConvergeTau_CompV0V1} (a) we thus characterize  the spatial  correlation length $\xi$ and the entanglement entropy $S$ of half of the ground state $\ket{\Omega}$. In this setting, $\tau$ only controls the accuracy of the Trotter expansion, and we thus expect the results to converge to a finite value as $\tau$ decreases. Once more both $S$ and $\xi$ stay finite at the critical point as a result of the finite bond dimension $\chi$. Our expectations are confirmed by the numerical results presented  in panel $(a_3)$ and $(a_4)$ of Fig. \ref{fig-ConvergeTau_CompV0V1}. 

In order to further verify that the tMPS is continuous, we compare its structure with the one of a cMPS. The continuous limit of an MPS was discussed in the context of the Bethe ansatz \cite{KM10cMPS01,MK10cMPS01}. In the tensor network community, the cMPS was first proposed by Verstraete \textit{et al} \cite{VC+11}. The cMPS can be used as the variational state for finding ground states of quantum field theories, as well as to describe real-time dynamical features.The cMPS  describes the low-energy states of quantum field theories once appropriately regularized, in the same way as a normal MPS describes the low-energy states of quantum spin systems. The cMPS can be constructed as the continuous limit of a discrete MPS defined as (see \cite{VC+11}),
\begin{equation}
\vert \psi \rangle = \sum_{n_1 \cdots n_L} V^{n_1} \cdots V^{n_M} \left( \Psi^{\dagger}_1 \right)^{\sigma_1} \cdots \left( \Psi^{\dagger}_M \right)^{n_M}\vert \Omega \rangle ,
\end{equation}
where the $V$s satisfy
\begin{align}
V^0 &= I - \tau Q \label{eq:cv0} \\
V^1 &= \tau R \label{eq:cv1} \\
V^n &= \tau^n R^n \label{eq:cv2}\\
\Psi &= \frac{\hat{a}_i}{\sqrt{\tau}} . \label{eq:cv3}
\end{align}
Both $R$ and $Q$ are independent from $\tau$.

We can thus use in particular \eqref{eq:cv0} and \eqref{eq:cv1} to extract $Q$ and $R$ for the Ising field theory, defined by the tMPS. In Fig. \ref{fig-ConvergeTau_CompV0V1} (b) we show the components $V^0$ and $V^1$ as a function of $\tau$. In the log-log plot we see that once we appropriately subtract the identity component to $V_0$, $V_0$ and $V_1$ scale differently  as predicted by  Eq. (\ref{eq:cv0}) and (\ref{eq:cv1}). 
This implies that by performing an appropriate scaling analysis, we can directly extract the $R$ and $Q$ for the tMPS of the Ising field theory.  We only need to extrapolate the results at finite $\tau$ to the interesting $\tau\to 0$ limit, thus overcoming the difficulties that arise when trying to directly optimize the continuous MPS \cite{VC+11, GRV+17, SHM+15}.

\section{Identifying the emerging quantum field theory}

\begin{figure}[t]
    \begin{center}  
	\includegraphics[width=\textwidth]{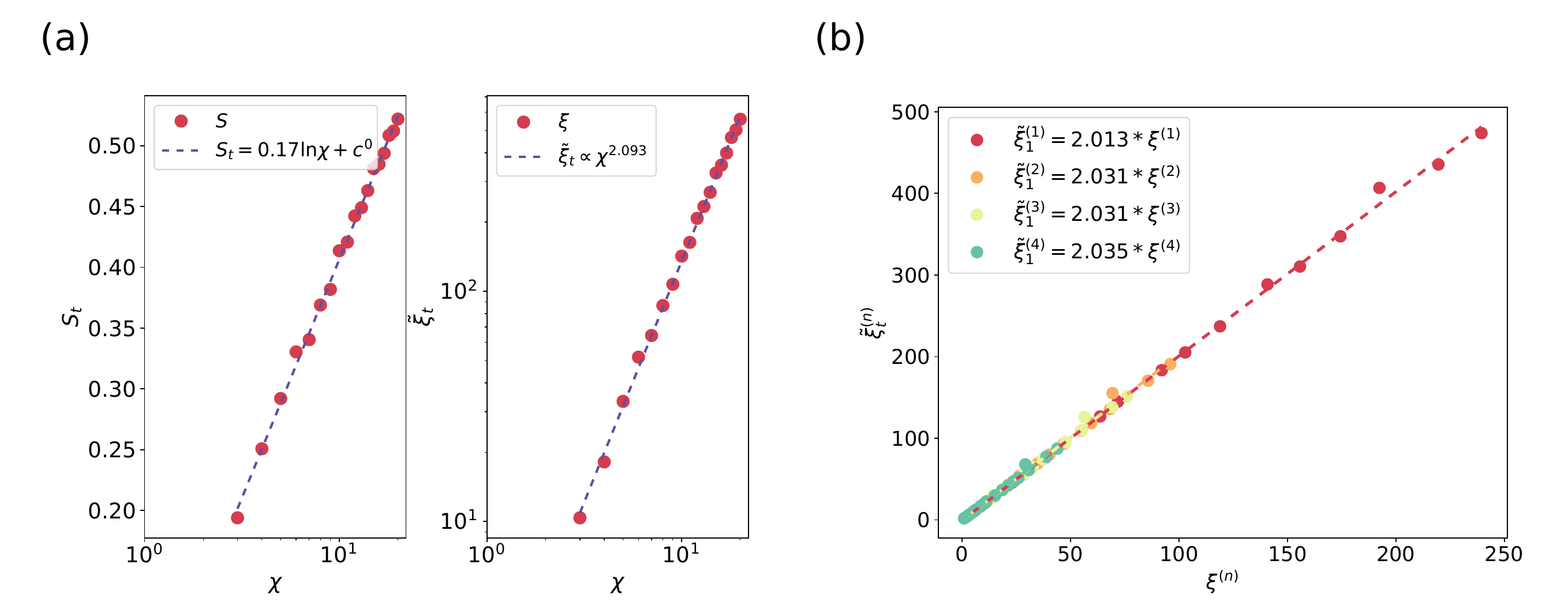} 
	\end{center}
	\caption{\textbf{Temporal entanglement entropy and correlation 
	Time.}  
	(a) The scaling of the correlation time $\tilde{\xi}_T$ 
	and the temporal entanglement entropy $S_T$ versus the bond     
	dimension cut-off $\chi$ of the tMPS. We take $h=0.5$, $L=2$ 
	and $\tau=10^{-6}$. (b) Correlation length of the tMPS 
	$\tilde{\xi}^{(n)}_T$ versus that of the spatial MPS 
	$\xi^{(n)}$ for different bond dimensions $\chi$ and 
	$\tau =10^{-6}$. Our results show that 
	$\tilde{\xi}^{(n)}_T = \nu \xi^{(n)}$ with $\nu=2$.}
	\label{fig-XiXiT}
\end{figure}

By taking the continuum limit at the critical point we expect to recover a critical field theory. However we already know that the 
finite $\chi$ of MPS induces a finite correlation length  \cite{TOI+08,PMT+09,SHM+15,LT+12} scaling as
\begin{eqnarray}
\xi &\propto& \chi^\kappa. \label{eq-Xiscaling}
\end{eqnarray}
By using the Cardy-Calabrese results close to the critical point, this implies that the  entanglement entropy $S$ should scale as 
\be \label{eq-Sscaling}
S = \frac{c}{6} \ln \xi + \mathit{c'}= \frac{c\kappa}{6} \ln \chi + \mathit{c'}
\ee

In order to characterize it, we use the CTMRG algorithm described in Refs. \cite{NO+96,NOHMA+00,NOHMA+01,NOK+96}. 
Our results for the  ground state MPS in Fig. \ref{fig-Q1D_spatial} (a) 
reproduce correctly Eq.  \eqref{eq-Sscaling}.  
In Fig. \ref{fig-Q1D_spatial} (b) we present the scaling of $S$ against 
$\xi$ of the spatial MPS that allows extracting the central charge, whose value turns out $c=0.502(5)$ as expected where the error is obtained by fitting several subsets of data for different $\chi$ and $L$.

Close to the critical point, at low energies, the system becomes not only scale-invariant but also Lorentz invariant, as a consequence of the linear dispersion relation for low-energy excitations. For this reason, it is possible to rotate the system and invert the role of space and time. In this way, we can think of the tMPS as a state  along the infinite temporal direction, and we expect it to share some properties (e.g. the scaling exponents) with the matrix-product state defined along the spatial direction.

We now characterize the correlation time $\tilde{\xi}_T$ and the entanglement entropy $S_T$ of the tMPS (that has been called temporal entanglement in Ref. \cite{HM+15}). As shown in Fig. \ref{fig-XiXiT} (a) $\tilde{\xi}_T$ and $S_T$ also satisfy Eqs. (\ref{eq-Xiscaling}) and (\ref{eq-Sscaling}),respectively. From our best fit, we extract $\kappa = 2.026(4)$ and $c_T=0.504(3)$ for the tMPS compatible with the ground state data. 

The anisotropic continuous limit introduced in the Trotter expansion, that only involves the temporal direction, induces a non-trivial dependence  between the physical quantities extracted from the MPS and the those extracted from the tMPS.
This can be understood to be the analog of the well-known physics of classical anisotropic models. They behave in the same way as their isotropic counterparts once non-universal rescaling factors are taken into account. For example, the $2D$  anisotropic Ising model, where the coupling $J_s$ and $J_t$ along the two directions are different, possesses a line of critical points, all in the Ising universality class.  
They can be found from the equation $\sinh{(2\beta J_t)}\sinh{(2\beta J_s)}=1$,  where $\beta$ is the inverse temperature (see, e.g. \cite{kogut+79}).
 
The rescaling factor $\nu$ can be extracted from the ratio of the correlation lengths along the two directions
\begin{eqnarray}
\xi = \nu^{-1} \tilde{\xi}_T, \label{eq-XiXiT}
\end{eqnarray}
that encodes the speed of sound.
We can extract $\nu$ from our data  in  Fig. \ref{fig-XiXiT} (b) and find it to be roughly $\nu=2$.
We verify that using the same speed we can transform all the gaps of the temporal transfer matrix into the corresponding ones of the space transfer matrix.

We define
\begin{align}
\tilde{\xi}^{(n)}_T &= \frac{1}{\tilde{\Delta}_n}= \left( \log \frac{\tilde{\eta}_0}{\tilde{\eta}_n} \right)^{-1} , \\
\xi^{(n)} &= \frac{1}{\Delta_n}= \left( \log \frac{\eta_0}{\eta_n} \right)^{-1},
\end{align}
where $\eta$ and $\tilde{\eta}$ are the low-lying eigenvalues of the spatial and temporal transfer matrix. Note for $n=1$, we have that 
$\tilde{\xi}^{(1)}_T=\tilde{\xi}_T$ and $\xi^{(1)}=\xi$. 

\begin{figure}[t]
     \begin{center}
	 \includegraphics[width=0.5\textwidth]{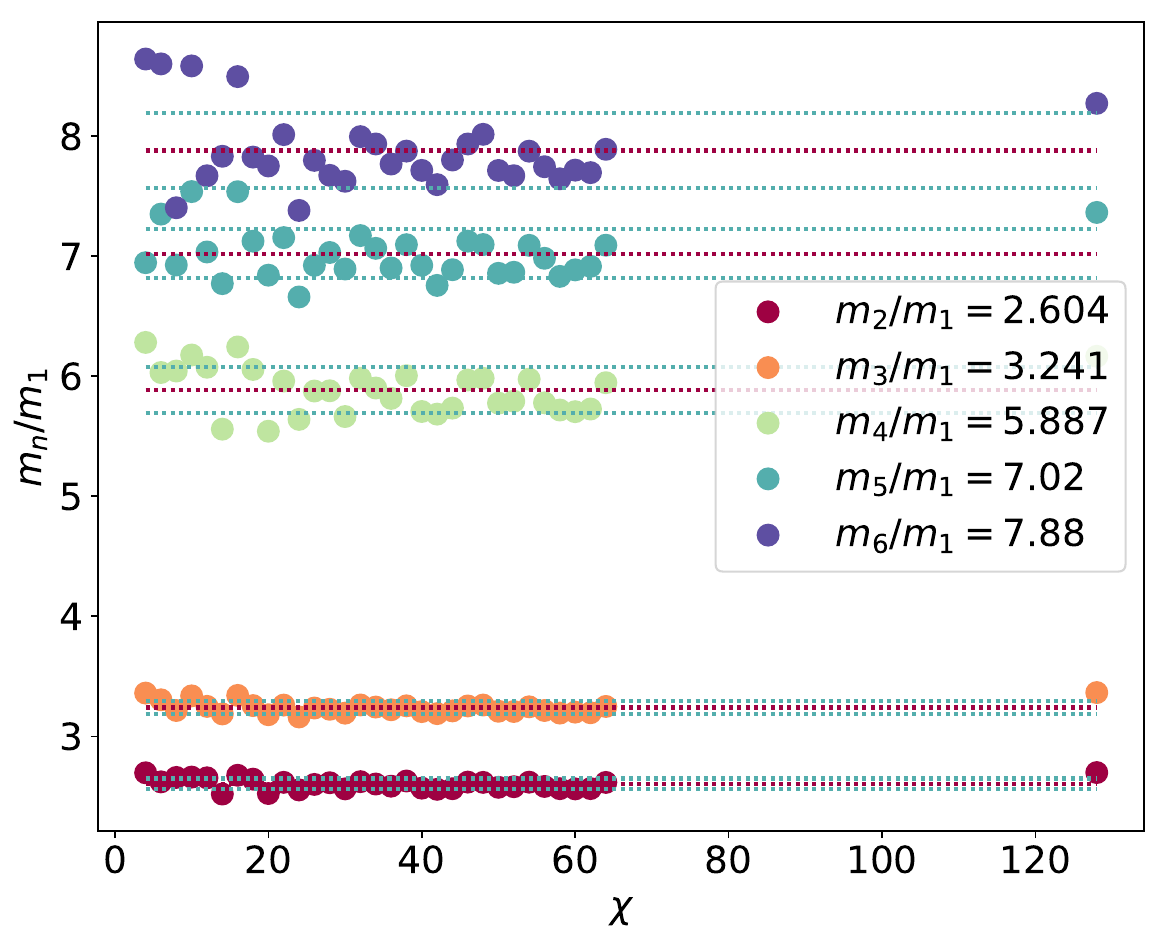} 
	 \end{center}
	\caption{\textbf{Spectrum of the tMPS transfer matrix.} 
	The ratios of the eigenvalues of the temporal transfer matrix 
	$E_T$ versus $\chi$ at $\tau =10^{-6}$. We use it in order to 
	identify the field theory that the tMPS describes.  	 
	\label{fig:ratios}}
\end{figure}

In the same Fig. \ref{fig-XiXiT} (b) we see that for all the  $\chi$ we have considered, the gaps of the temporal transfer matrix are indeed obtained from the gaps of the spatial transfer matrix by multiplying by the same speed $\nu=2$. This fact confirms our interpretation of $\nu$ as the non-universal velocity of the excitations.

The tMPS is continuous but its  entropy is  finite. The bond dimension of the tMPS indeed  plays the role of a IR regulator, meaning that we can think of the tMPS as describing a state that after coarsegraining a finite number of steps becomes a product state \cite{TOI+08}.
The field theory emerging in the continuum limit cannot be critical but must be massive, where the mass  is generated by the presence of a finite bond dimension closed to the Ising fixed point.

Given the emerging low-energy Lorentz invariance, we can identify the field theory spectrum by using the Bisognano Wichmann theorem.
The 1+1D version of the theorem states that the reduced density matrix of an half-infinite chain is related to the generators of boost, $\rho_{\textrm{half chain}} = \exp - K$ with $K=\int dx x T_{00}(x)$, with $T_{00}$ the time-time component of the stress-energy tensor \cite{bisognano_duality_1975}. In a CFT, the half chain can be mapped to a strip through a conformal transformation, 
thus proving that the spectrum of the reduced density matrix and that of the field theory on a strip are equivalent \cite{cardy_entanglement_2016,cho_universal_2017}.

\begin{figure}[!htb]
 \begin{center}
	 \includegraphics[width=0.35\textwidth]{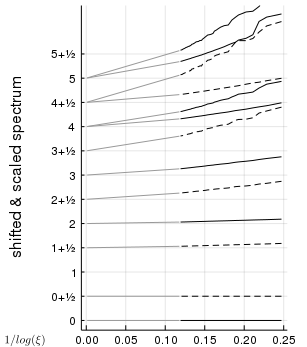} 
	 \end{center}
	 \caption{\label{fig:ent_spectrum} {\bf The emergence of the Ising field theory spectrum}. Entanglement spectrum as a function of the inverse logarithm of the correlation length. The data are presented with solid black lines, for even states under $Z_2$ sector and dashed lines for the odd states. The grey lines, are used as a guide to the eye in order to identify the closest CFT predictions. At leading order indeed the data should approach the CFT predictions as  $1/log(\xi)$ (hence the scale on the horizontal axis). In the plot we appreciate taht the first eigenvalues correctly reproduce the CFT spectrum. We also appreciate that the finite correlation length corrections become larger and larger as we increase the (entanglement) energy. This fact is a manifestation of the cross-over from a Lorentz invariant field theory at low energies to non-Lorentz invariant field theory described by the tMPS.}
    \end{figure}

Our conjecture that the tMPS represents the wavefunction of the ground state of the field theory thus can be checked by comparing the spectrum of the tMPS half-system reduced density matrix with what expected from the massive field theory obtained by deforming a CFT \cite{cho_universal_2017}. 

We thus analyse the entanglement spectrum of the tMPS. The data are presented in Fig. \ref{fig:ent_spectrum}. There we compare the results extracted from cMPS with $Z_2$ symmetry to the one expected from the Ising CFT with free boundary conditions, that include the tower of the identity operator, equispaced starting from $0$, and the tower of the $\sigma$ operator, equispaced starting from $0.5$.
In the identity tower the descendant at $1$ is absent. 

The numerical data are shifted and rescaled in such a way that the first energy is at $0$ and the first excitation is at $0.5$ (following the analysis on the lattice performed by Lauchli in \cite{lauchli_operator_2013}). The spectrum is plotted as a function of the inverse of the logarithm of the finite correlation length induced by the finite bond dimension, since this is the variable dictating how the gaps in the spectrum should close.

We appreciate that the tMPS correctly describes the field theory at low energies. The CFT predictions are on the y axis. We compute the spectrum using $Z_2$ symmetric tensors and solid lines represent the $Z_2$ even states while dotted lines the $Z_2$ odd states. The grey lines are presented as guide to the eyes, and connect the numerical values to the theoretical one, linearly as expected by the leading closure of the gap as $1/log(\xi)$.  We appreciate that at low energies, the tMPS spectrum coincides with the one of the Ising CFT on a strip with free boundary conditions on both edges. 

The tMPS spectrum does however systematically deviates from the field theory one at higher energies. When the same study is performed on the lattice, the deviation is expected to be caused by lattice artefacts \cite{lauchli_operator_2013,hu_emergent_2020}. Here however we are in the continuum and we thus need to interpret the deviations in a different way. We believe that the deviation need to be understood by the fact that a finite-bond dimension CMPS cannot encode a Lorentz invariant state at short distances (high-energies), where the entanglement entropy should diverge.

The CMPS state can however encode a state that crosses-over from Lorentz invariant at low energies to non Lorentz invariant at high energies. The deviation of the continuous tMPS spectrum at high energies from the one expected from the CFT should thus be a manifestation of such cross-over  \footnote{LT thanks Guifre Vidal for in depth discussions and clarifications on the topic}.

\section{Conclusion}

We have explicitly constructed the ground state of a QFT obtained from the continuous limit of the Ising model as an MPS with fixed bond-dimension. We have done it by exploiting both Hastings's observation that the Euclidean time evolution generates a continuous MPS in the imaginary time direction and the low energy Lorentz invariance emerging at the critical point. 

These two ingredients allow characterizing the QFT by studying 
the properties of the tMPS, a continuous MPS emerging, as a result of the Trotter expansion, along the temporal direction. Having obtained a field theory in the vicinity of the Ising critical point a natural question is if we can obtain its spectrum by the simple modification of the Ising fixed point. 

In order to answer this question, we have characterised the spectrum of the reduced density matrix of half-infinite tMPS and found it in agreement at low (entanglement) energies with the one of the Ising field theory perturbed by a ``thermal'' relevant operator \cite{cho_universal_2017}.   

Our results seem to suggest that the finite bond dimension suffice in order to have a well defined field theory in the continuum, without any further cut-off.  This aspect needs to be further characterised and 
understood, as possible generalization of our construction to higher dimensions. 

\section*{Acknowledgements}
E.T. and L.T. acknowledge the discussion with Giuseppe Mussardo, Pasquale Calabrese on the mass spectrum of the perturbed Ising model and Andreas L\"{a}uchli and Philippe Corboz for discussions on related subjects.  S.J.R. acknowledges Fundaci\'o Catalunya - La Pedrera $\cdot$ Ignacio Cirac Program Chair and is supported by Beijing Natural Science Foundation (No. 1192005 and No. Z180013), Foundation of Beijing Education Committees (No.
KM202010028013), and the Academy for Multidisciplinary Studies, Capital Normal University.
We acknowledge the Spanish Ministry MINECO (National Plan
15 Grant: FISICATEAMO No. FIS2016-79508-P, SEVERO OCHOA No. SEV-2015-0522, FPI), European Social Fund, Fundació Cellex, Fundació Mir-Puig, Generalitat de Catalunya (AGAUR Grant No. 2017 SGR 1341, CERCA/Program), ERC AdG NOQIA, EU FEDER, MINECO-EU QUANTERA MAQS, and the National Science Centre, Poland-Symfonia Grant No. 2016/20/W/ST4/00314. 
\begin{appendix}

\newpage

\section{Two-dimensional classical Ising partition function} \label{2DCIM}

\begin{figure}[t]
    \begin{center}
	\includegraphics[scale=0.45]{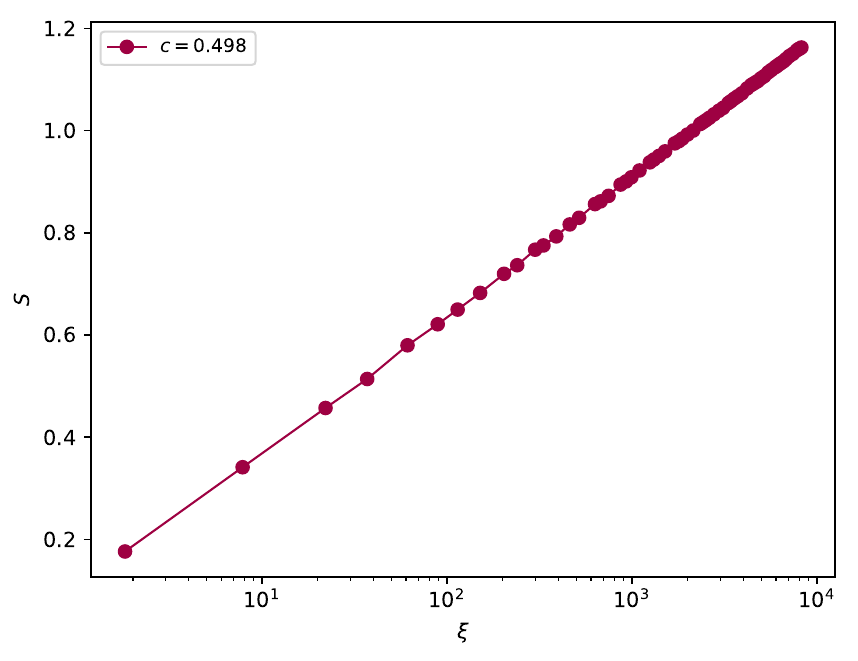}
	\end{center}
	\caption{\textbf{Temporal entanglement entropy vs correlation length 
	.} The temporal entanglement entropy $S$ as a function 
	of the correlation length $\xi$ at the critical temperature. Each
	point is obtained by choosing the bond dimension $\chi = [2,1280]$. 
	A 	fit to the expected logarithmic scaling allows to extract $c = 
	0.5$ as expected. The results from the spatial MPS are the same 
	as those from the tMPS with the difference $O(10^{-5})$.}
	\label{fig-Ising2D}
\end{figure}

In this section, we apply the TN encoding scheme to 2D classical Ising model at the criticality. There are some intrinsic differences between this TN of 2D classical partition and that of the (1+1)D quantum theory. For the 2D classical partition, the two dimensions of the TN are both spatial and discrete, and they are equivalent to each other.

Its partition function can be directly written in a 2D TN, where the local tensor is the probability distribution of some local Ising spins. Here, we take the Ising model on square lattice as an example, where the local tensor is defined as
\begin{eqnarray}
T_{s_1s_2s_3s_4} = e^{-\beta (s_1s_2+s_2s_3+s_3s_4+s_4s_1)},
\end{eqnarray}
with $\beta$ the inverse temperature and the spin index $s_i=\pm 1$. Note that the local tensor of the TN can also be chosen as the contraction of several $T$'s.

It is well-known that the dominant eigenstate of the transfer matrix can be approximated as an MPS. Each MPS with a finite bond dimension $\chi$ 
corresponds to a gapped state with a finite correlation length $\xi$ and 
entanglement entropy $S$. At the critical temperature, the central charge can be extracted by the scaling behavior of $S$ against $\xi$ 
\cite{VLR+03,TOI+08,PMT+09}. Specifically speaking, with different $\chi$, $\xi$ and $S$ satisfy
\begin{eqnarray}
S =\frac{c}{6} \ln \xi + \mathit{const}, \label{eq-Sxi}
\end{eqnarray}
where the coefficient gives the central charge $c$. Note that Eq. 
(\ref{eq-Sxi}) is independent of calculation parameters. Meanwhile, one can also check the scaling behavior with different $\chi$'s separately and have
\begin{eqnarray}
\xi &\propto& \chi^\kappa, \label{eq-xidc} \\
S &=& \frac{c\kappa}{6} \ln \chi + \mathit{const} . \label{eq-Sdc}
\end{eqnarray}
By substituting, one can readily have Eq. (\ref{eq-Sxi}) from these two 
equations. For the 2D Ising model, we have the critical temperature 
$\beta_c=\ln(1+\sqrt{2})/2$ from the exact solution \cite{O44} and $c=1/2$ that corresponds to a free fermionic field theory \cite{VLR+03}. We shall stress that for any finite $\chi$, we cannot exactly give a critical state by MPS, but only a gapped. The central idea of the scaling theory with MPS is to extract the conformal data from the scaling behaviors the gapped MPSs.

In Fig. \ref{fig-Ising2D}, we show that the spatial MPS has the same 
correlation length $\xi$ and entanglement entropy $S$ as the temporal MPS with the difference $O(10^{-5})$. In other words, these two MPSs, though updated within two different schemes and located in two different directions of the TN, are connected by a gauge transformation. By choosing different bond dimension cut-offs and cells to construct the tensor $T$, the relation between $\xi$ and $S$ shows a robust logarithmic scaling, giving accurately the central charge.
\section{The tMPS transfer matrix puzzle}
In previous version of this manuscript we tried to identify the field theory encoded by the tMPS by characterising its transfer matrix. 
By using the Bisognano Wichmann theorem we now have been able to identify the field theory with that of a massive deformation of the Ising fixed point, but for coeherence with our previous attempt we want to report some of the checks we have performed on the transfer matrix spectrum.
In our previous version we have used the possibly too simplistic assumption that the  the low energy spectrum of the field theory $m_1,m_2,m_3,m_4$ can be extracted by constructing the appropriate ratios of the tMPS transfer matrix eigenvalues, namely $m_n= 1/\tilde{\xi}^{(n)}_T$.  The identification of the gaps of the transfer matrix with the low energy masses of the field theory. This identification was partially justified in \cite{zauner_transfer_2015}, for the low lying real eigenvalue of the transfer matrix. 
The results we present in this section show that at least the naive identification cannot hold for the higher eigenstates of the tMPS

The correct way of extracting the operator content is to  analyse the zero momentum sector of the retarded two point correlation functions. At sufficiently large times, the correlator can indeed  be expressed as a series of decaying exponential one per masses in the field theory \cite{Cardy+861,banuls_spin_2020}. 

In \cite{zauner_transfer_2015}, the identification of zero momentum eigenstate with the eigenvectors of the MPS transfer matrix with real eigenvalues was pushed forward. However by working on the lattice, it is not clear at all that the real eigenvalues correspond only to the zero momentum eigenstate, and thus it is not clear that without an appropriate zero momentum projection, the eigenvalues of the tMPS transfer matrix bear a relation to the field theory spectrum. 

In order to clarify this statement below  check if, by using the real and symmetric gauge for the Ising tMPS, where all the eigenvalues of the transfer matrix are real, we can identify the spectrum of the transfer matrix with any of the spectra of field theories obtained by deforming the Ising fixed point. 

After long discussion we are now aware that in principle all momentum sectors of the retarded two point correlation function could mix, and we have been unable to devise an easy way to project onto the zero momentum sector. However, we still think that it is important to report the systematic analysis we have perform when we assumed that the lack of a proper zero-momentum projection should not affect our capability of identifying the low energy spectrum encoded in the tMPS transfer matrix.  

The masses extracted by the full tMPS spectrum, presented in Fig. \ref{fig:ratios}, are in good agreement with those reported elsewhere \cite{SHM+15}. Since we ignore what are the boundary conditions induced by the infinite MPS, we start to compare, first, the masses of the tMPS with the masses of the quantum Ising model at criticality in the presence of single impurity, that should allow to describe all possibile boundary conditions.

The fact that the masses depend on the strength of the impurity was discussed in detail in 
\cite{Cardy+861,barev,mccoy_1980,oshikawa_1997,EV2014}. We consider a quantum critical system with Hamiltonian of the form
\be 
\hat{H}^{\rm Imp}= \hat{H}+J^{\rm Imp}_R ,
\ee 
where $\hat{H}$ is a fixed-point Hamiltonian that describes the host system, and $J^{\rm Imp}_R$ accounts for an impurity localized on a region $\mathcal{R}$ of the lattice. Specifically, we test our results in the case where $\hat{H}$ corresponds to the critical Ising Hamiltonian defined in eq. (\ref{eq-IsingHam}) and the impurity Hamiltonian $\hat{J}^{\alpha}$ acts on two adjacent lattice sites $r \in (0,1)$ where it weakens or strengthens the nearest neighbor term,
\be 
J^{\alpha}(0,1)= \left(1-\alpha \right) \sigma^x_0 \sigma^x_1
\ee
for some real number $\alpha$. The quantum critical Ising model with an impurity of this form, which is direct correspondence with $2D$ classical Ising model with defect line, has been extensively studied in the literature \cite{EV2014}. 
For each value of impurity coupling $\alpha$  we also compute the scaling dimensions $\Delta_{\alpha}$ associated to the impurity.
In Ref. \cite{EV2014} the spectrum of scaling dimensions for the critical Ising model associated to the impurity $J^{\alpha}$ have been derived analytically
\be 
\Delta_{\alpha}=2\left(c+\frac{1}{4}+\frac{1}{\pi} \tan^{-1}\left( \frac{\alpha}{1+\alpha}\right) \right)^2 +m
\ee 

In \cite{bal_2016} it was indeed proposed that the physics of the MPS should indeed be dictated by that of an impurity in the path integral representation of the partition function. Encoding such partition function as the scalar product of two MPS states implicitly introduces a line of impurities located where the ``physical'' legs of the MPS are. We are thus tempted to identify the strength of such impurity by matching the tMPS spectrum with the one that comes from the impurity.

\begin{figure}[t]
     \begin{center}
	 \includegraphics[width=0.5\textwidth]{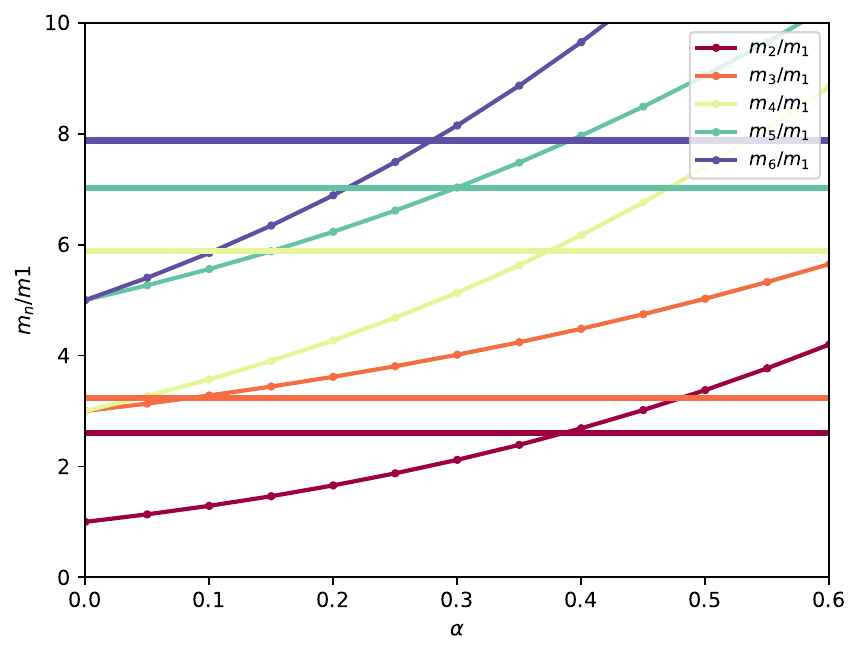} 
	 \end{center}
	\caption{\textbf{Ratios masses $m_n/m_1$ for the critical 
	Ising model with a conformal defect $J^{\alpha}$.} 
    Comparison between numerical results from the tMPS (solid 
    lines) with the analytical results (dotted lines).  
    Note that only scaling dimensions in the $p=-1$ parity sector 
    of the $Z_2$ global symmetry of the Ising model are 
    plotted, as those in the $p=+1$ parity sectors are invariant 
    under addition of the conformal defect.}
	\label{fig-ratio_imp}
\end{figure}

In order to compare the value of  the numerical masses $m_n$ we obtain, we need to get rid of the arbitrary rescaling factor in front of them. For this reason we compare the ratios of the masses in Fig. \ref{fig:ratios}.

In Fig. \ref{fig-ratio_imp}, we compare the ratios for small impurity strength $\alpha$ with those we obtain numerically for the tMPS. We see that there is no single impurity strength that allows to match our tMPS spectrum at low impurity strength. At impurity around $1$, we see also no possible matching between the two spectra since $m_{2}/m_{1}$ is already $8$. Now if we  further increase the impurity above one, the first ratio increases from $8$ to infinitely since $1$ is fixed while the scaling dimension of the spin operator goes to zero. 
Our data thus exclude that without zero momentum projection on the transfer matrix spectrum the simple impurity scenario proposed in \cite{bal_2016} can give account of the transfer matrix spectrum we observe.

\begin{figure}[t]
     \begin{center}
	 \includegraphics[width=1.0\textwidth]{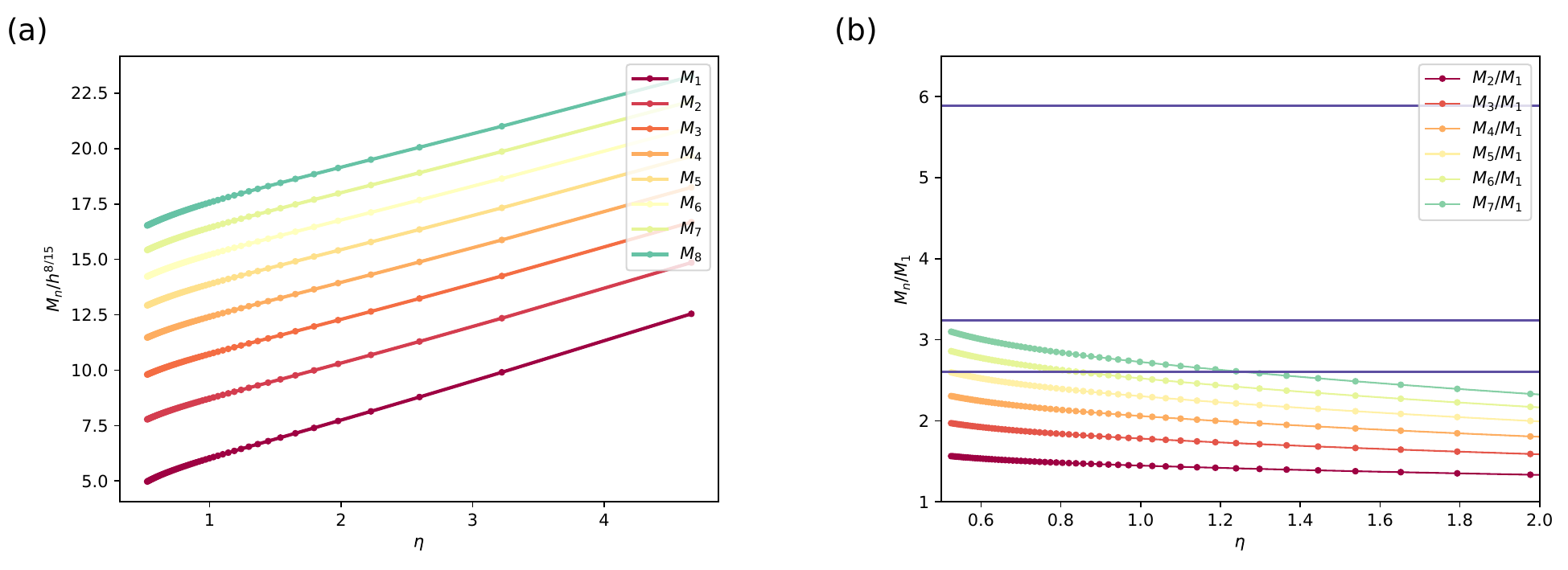} 
	 \end{center}
	\caption{\textbf{Ratios of masses of quantum field theory.} 
	(a) The mass spectrum $M_n(\eta)$ of Ising field theory 
	particles at positive $\eta$ a obtained using the WKB 
	approximation. The present results reproduce  those in Ref. 
	\cite{Zam-89}, thus allowing to identify the field theory 
	encoded in the tMPS with the one obtained 
	by weakly perturbing an Ising model with a magnetic field. 
    (b) The ratios of 	masses of Ising field theory at positive $
    \eta$. 
    The solid blue lines represents the masses ratios obtained 
    numerically from temporal transfer matrix $E_T$ at $\tau 
    =10^{-6}$ and $\chi=128$.  	 
	\label{fig:ratios_comp}}
\end{figure}

We now turn to possible modification of the Ising field theory with relevant operators. 
As regards the Ising field theory, the model, which arises as to the continuum limit of the quantum Ising chain, is described by the Hamiltonian
\be 
\hat{H}= \int dx \left[ i \left(\overline{\Psi} \partial_x \overline{\Psi}-\Psi \partial_x \Psi +m \overline{\Psi} \Psi \right)  +g \sigma \right]
\ee
where $\Psi$ ($\overline{\Psi}$) is the left (right) moving Majorana fermion field, $m$ is the fermion mass related to the transverse magnetic field in the lattice model, and $g$ is equivalent to a longitudinal magnetic field in the lattice model \cite{fonseca_2006}. Here we will closely follow that analysis in order to understand if the spectrum of the tMPS transfer matrix arise from some of those deformations of the Ising field theory. The physics of Ising field theory essentially depends on a single dimensionless scaling parameter, which we define as
\be 
\eta=\frac{m}{h^{8/15}} .
\ee 
In the ordered phase, the fermions can be regarded as the product of domain wall excitations and spin flips in the ferromagnetic spin configuration. The
longitudinal field $g$ acts as a non-local confining potential
for the domain walls. 
While the field theory is non-integrable for generic values of the parameters $m$ and $g$, there exist two special lines in parameter space ($m = 0$ and $g=0$) along which the model is integrable.

In the ordered phase of the model, corresponding to $m>0$, 
the spectrum of the model depends intimately on the longitudinal field: 
when $g=0$ a flipped spin fractionalizes into two independent domain walls, which can freely propagate through the system. Thus at low energies, when $g=0$, there is a two-particle continuum of excitations, separated from the ground state by an energy gap of $2 m$. 
On the other hand, when $g\neq0$ there are profound changes in the spectrum. The presence of the longitudinal field, which is nonlocal in terms of the domain wall fermions, induces a linear potential between domain wall excitations, leading to confinement. This is very
much reminiscent of the formation of mesons in quantum chromodynamics. The low energy spectrum now completely restructures the two domain wall continuum at $g=0$ collapses into well-defined meson excitations for $g\neq0$, with a new multimeson continuum forming above energies $E=4m$.
For weak longitudinal field $g$ and away from the two-particle threshold $2m$ the energy of a meson state can be expressed as
\be \label{eq:ZamoEq}
M_n = E_0 + 2m \cosh\left(\Theta_n \right) \qquad n=1,2,\ldots ,
\ee
where $E_0$ is the ground state energy and $\Theta_n$ is a rapidity that satisfies the non-linear quantization condition
\be 
\sinh \left(\Theta_n \right) -2 \Theta_n = 2\pi \lambda \left(n-\frac{1}{4} \right) -\lambda^2 S_1\left(\Theta_n\right) -O(\lambda^3),
\ee
where $\lambda=2 \overline{\sigma} g/m^2$ with $\overline{\sigma}=|m|^{1/8} \overline{s} $ and $\overline{s}=2^{1/2} e^{-1/8} \mathcal{A}$ with $\mathcal{A}=1.282427 \ldots$ being Gleshier's constant. We have also defined the function
\be 
S_1\left(\Theta_n\right) = \frac{-1}{\sinh\left(2 \Theta_n \right)} 
\left[\frac{5}{24} \frac{1}{\sinh^2\left(\Theta_n \right)} +\frac{1}{4} \frac{1}{\cosh^2\left(\Theta_n \right)}-\frac{1}{12}-\frac{1}{6} \sinh^2\left(\Theta_n \right)  \right]
\ee
The spectrum of mesons presented in Fig \ref{fig:ratios_comp} (a) is dense in at large $\eta$. As $\eta$ decreases, the heavier mesons become unstable against decay into the lighter ones, and successively disappear from the spectrum of stable particles, typically becoming resonances. 
    
In Fig. \ref{fig:ratios_comp} (b) we compare directly the numerical ratios $m_n/m_1$ with the analytically ones $M_n/M_1$ extracted from the Eq. (\ref{eq:ZamoEq}). We see that there is no single value of $\eta$ that allows fixing the ratios and Zamolodchikov's ratio are systematically lower.   
 
Summarizing, by working in the gauge where the tMPS gives rise to an Hermitian transfer matrix, all the transfer matrix eigenvalues are real. We expect that the zero momentum eigenstates of that TM should encode the masses of the corresponding field theory since they are the one that mediate the exponential decay of the corresponding retarded correlation in time. However by working with the tMPS where all the lattice sites have been coarse-grained to two effective sites with Hilbert space of dimension $\chi$ we  loose the information about the original momentum on the lattice.  

It is pretty well established that at least the lowest mass of the MPS transfer matrix should correspond to the lowest mass of the field theory. However, here the lowest mass is needed in order to match the scales of the tMPS with those of the corresponding field theory. 

By ignoring the lack of momentum resolution, we have tried to match the rest of the tMPS spectrum with the spectra of possible field theories constructed in the vicinity of the critical point but none of those match. 
This evidences that we still need to understand better what is the relation of the tMPS spectrum with field theory data.

\end{appendix}


\end{document}